# Coherent Signal Amplification in Bistable Nanomechanical Oscillators by Stochastic Resonance

R. L. Badzey and P. Mohanty

*Department of Physics, Boston University, 590 Commonwealth Avenue, Boston, MA 02215, USA*

**Stochastic resonance is a counter-intuitive concept[1,2]; the addition of noise to a noisy system induces coherent amplification of its response. First suggested as a mechanism for the cyclic recurrence of ice ages, stochastic resonance has been seen in a wide variety of macroscopic physical systems: bistable ring lasers[3], SQUIDs[4,5], magnetoelastic ribbons[6], and neurophysiological systems such as the receptors in crickets[7] and crayfish[8]. Although it is fundamentally important as a mechanism of coherent signal amplification, stochastic resonance is yet to be observed in nanoscale systems. Here we report the observation of stochastic resonance in bistable nanomechanical silicon oscillators, which can play an important role in the realization of controllable high-speed nanomechanical memory cells. Our nanomechanical systems were excited into a dynamic bistable state and modulated in order to induce controllable switching; the addition of white noise showed a marked amplification of the signal strength. Stochastic resonance in nanomechanical systems paves the way for exploring macroscopic quantum coherence and tunneling, and controlling nanoscale quantum systems for their eventual use as robust quantum logic devices.**

The conditions for a system to exhibit stochastic resonance are well known and fairly robust. First among these is an energy threshold. Second, there needs to be an applied modulation, which causes the system to cross the threshold. And finally, there needs to be a source of noise. One of the classic examples of a system which undergoes stochastic resonance is that of a particle in a double-well potential. With no modulation, the particle will transition between the two wells with the Kramers rate, $\Gamma_K \propto e^{-\frac{\Delta V}{D}}$, where $\Delta V$ is the depth of each well and D is strength of temperature-induced noise. With the addition of external periodic modulation, there is an interaction between the modulation and the noise, resulting in a resonance in the signal-to-noise ratio (SNR). The SNR is defined[2] as

$$SNR = 2\left[\lim_{\Delta\omega\to 0}\int_{\Omega-\Delta\omega}^{\Omega+\Delta\omega} S(\omega)d\omega\right]/N(\Omega) = \frac{S(\Omega)}{N(\Omega)} \quad (1)$$

where $S(\Omega)$ is the height of the power spectrum peak at the modulation frequency $\Omega$ and $N(\Omega)$ is the spectral background.

It is well known that with suitably strong driving the linear response of a damped, driven harmonic oscillator transitions to that of a nonlinear bistable system, which is the famous Euler instability. The equation of motion is described by the Duffing equation with driving force $F$ and driving amplitude $\omega$:

$$\ddot{x} + \gamma\dot{x} + kx \pm k_3 x^3 = F\cos\omega t. \quad (2)$$

Here, $k$ and $k_3$ are the linear and nonlinear spring constants, respectively, and $\gamma$ is the damping coefficient describing the dissipation of the oscillator. The sign of the $k_3$ term changes depending on whether the oscillator is under compressive (positive $k_3$) or tensile (negative $k_3$) strain. The presence of the cubic term in the equation of motion naturally leads to a term of fourth order in the beam potential,imparting that the oscillator now demonstrates bistability[9,10,11]. It is important to note regardless of the sign of the nonlinear term, both instances will result in a bistable oscillator. This is clear from the amplitude response of the Duffing oscillator: when looking at the amplitude response as a function of frequency, one finds that the oscillator no longer displays the symmetric Lorentzian lineshape characteristic of a damped, driven, linear harmonic oscillator. Rather, there is a sharp bend in the peak which creates a region of frequency space in which the oscillator is multi-valued. If the oscillator is softening, the bend occurs to the left of the location of the linear peak; likewise the bend is on the right in the case of a hardening spring. The signature of a system in nonlinear response is a hysteresis defined by sweeping the driving frequency forward and backward through the bistable region. For instance, in the compressive case, as the frequency is increased, the response will follow state 1 until reaching a point of instability, whereupon it will sharply drop to state 2. A downward sweep will produce the opposite effect, defining a region in which the beam is bistable; these situations are reversed in the case of a softening spring. Exciting at a single frequency within this region allows one to access either of the two bistable states by the addition of a suitable modulation[12]. In each case, bistable behavior only holds for driving frequencies near resonance.

Strictly speaking, the above description is that of a point particle in a double-well potential with no additional modulation or noise terms. In the case of a spatially-extended system such as a doubly-clamped beam or string under additional modulation and external noise, the more appropriate description is the Landau-Ginzburg equation, where the field variable $\Phi(x)$ replaces the position variable x:

$$\frac{d\Phi(x)}{dt} = m\Phi(x) - \Phi^3(x) + \kappa\frac{d^2\Phi(x)}{dx^2} + F_{mod}\cos\Omega t + \eta(x,t) \qquad (3)$$

Here, $\Omega$ is the modulation frequency and $\eta(x,t)$ is the added noise term. In their seminal paper, Benzi et al.[13] predict the emergence of stochastic resonance in a doubly clamped beam or string described by the Landau-Ginzburg equation with Von Neumann boundary conditions, $d\Phi(0)/dx = d\Phi(L)/dx = 0$. In addition to the two stable homogeneous solutions corresponding to the occupation of either well, there also exists a class of instanton solutions which imply that in the presence of noise the string can transition between the two wells. In this case, the string collective coordinate $u = (1/L)\int_0^L \Phi(x)dx$ synchronizes with the modulation within a certain range of noise powers. Therefore, the behavior of the collective coordinate as a particle in a double-well potential completes the projection of the doubly-clamped beam into the canonical single-particle picture described above.

We have fabricated two doubly-clamped nanomechanical beams from single-crystal Si with e-beam lithography and dry etching. Frequency sweeps with increasing drive amplitude show a marked evolution from the familiar Lorentzian response into a shifted response with a sharp drop and marked hysteresis. We perform forward and reverse frequency sweeps with large driving amplitudes to confirm and record the onset of nonlinear and bistable behavior. We then drive each bridge at a single frequency within its unique bistable frequency region in order to access both bistable states via the addition of a modulation signal of suitable strength. Our samples are driven with the well-known magnetomotive technique[14]; Figure 1 shows a schematic of the circuit used to drive the bistable oscillator and add in white noise.

Beam 1 is cooled to 300 mK and driven nonlinear with $A_{drive}$ = 4.0 dBm. The modulation is made too weak (-5.5 dBm) to induce switching, and the noise is swept from -71 dBm (~ 79 pW) to -41 dBm (~ 79 nW). At each noise power, a time series is taken to monitor the oscillator response. Figure 2a shows selected time series at different noise powers, with their associated power spectra. After calculating the SNR, the data are combined and represented in Figure 2b. The resonance is clear and dramatic, with a sharp increase in SNR around 18 nW. It is relevant to note that these noise powers are those coming from the source rather than the actual power incident on the sample.

It has been shown[15] that the temperature of the beam is a viable source of noise, causing a decrease in switching fidelity and increase in state noise at temperatures below 1K. Therefore, there should be a temperature or range of temperatures above 1K where we would actually see an *increase* in the signal-to-noise ratio. Beam 2 is disconnected from the noise source. In this case, the beam achieves nonlinear response with a smaller driving amplitude (1.0 dBm) while the amplitude necessary for switching is much greater (> 16.5 dBm). Because the modulation power is so large, the electrical signal actually shows up in our lock-in technique, as can be seen in Figure 3a. Again, the left-hand side shows segments from the time series at each temperature (with the modulation amplitude highlighted in gray), while the right-hand side shows the power spectra obtained by FFT. It is immediately apparent that the switching is much noisier than that of beam 1, even when one takes into account the background effect of the modulation. Full switching is never really achieved, even for the most responsive time series.

The difficulty in inducing switching behavior can be understood by considering the symmetry of the double-well potential. The response of beam 1 lent itself to the conclusion that the potential is very nearly symmetric. However, it rapidly became apparent that such was not the case for beam 2. This is especially evident when looking at Figure 3a, Graph B. At each impulse from the modulation, there is a sharp spike up to the upper state, which immediately decays down to the lower state; this is an indication that the potential of the bridge is asymmetric. The occurrence of incoherent switches down to the lower state is so prevalent that even when the SNR is maximized, the switching is not truly coherent. This is in sharp contrast to the very clean, symmetric behavior shown by beam 1. In addition, a look at the power spectra from bridge 2 reveals the presence of even-order harmonics, a sure signature of an asymmetric potential[16, 17]. It is important to mention here that the spring coefficients $k_1$ and $k_3$ change with temperature, which might affect both the symmetry and even the shape of the potential. Such a change is manifested by a shift of the resonance frequency of the beam with a change in temperature, which is proportional to the sound velocity $v = \sqrt{E/\rho}$, where $E$ is the Young's modulus and $\rho$ is the material density. As the temperature is reduced, the modulus is increased, leading to a subsequent increase in the resonance frequency. However, we found that such frequency shifts were on the order of less than $10^{-4}$ within the temperature range we employed during these experiments.

Another important question is the effect of temperature on the dissipation coefficient $\gamma$. Should the effect be non-negligible, then there exists a multiplicative noise term in the Duffing equation. This is in sharp contrast to the purely additive noise of the classic theory of stochastic resonance. The dissipation is most easily quantified by measuring the quality factor $Q$ of the oscillator as a function of temperature. Our measurements of this quantity indicate that $Q$ can change by as much as 10% over the temperature range from 2 – 3 K. While this is not a huge effect, it can certainly have an impact on the system noise, and is definitely worth exploring in more detail. The topic of stochastic resonance has been broached in the literature[18,19], but this provides an avenue for continued research (A. Bulsara, private communication).

Again, an analysis of each power spectrum produces the SNR value, and these points are compiled to create Figure 3b. The peak of SNR is much sharper than that seen in beam 1. This might at least in part be due to the rather coarse sweep, when looked at from the perspective of input noise. The apparent increase in SNR at higher temperatures (> 3000 mK) is somewhat misleading. This is due to the still-present signal from the modulation which produces a peak in the power spectrum and is not due to actual switching between the two bistable states. Filtering the data to excise the effect of modulation would undoubtedly produce a cleaner power spectrum, but the data is presented raw in order to underscore the presence of SR in a system that is less than optimal in potential symmetry and background noise.

In sharp contrast to most other systems investigated for stochastic resonance we use a square wave modulation, as opposed to the canonical sine wave. This is an experimental consideration, as we found switching was much more easily achieved with a square wave. While a seemingly minor technical difference, there are some important ramifications brought on by the use of this type of modulation. First, the modulation signal can be decomposed into the form

$$S_{\text{mod}} = \sum_{n=1}^{\infty} a_n \cos n\omega_{\text{mod}} t + b_n \sin n\omega_{\text{mod}} t . \tag{3}$$

Here $S_{\text{mod}}$ is the input signal, $n$ is the harmonic number, and $\omega_{\text{mod}}$ is the modulation angular frequency. For a linear response system which follows McNamara *et al.*[20] this does not significantly alter the SNR behavior; the ratio is still dominated by the fundamental frequency. But under the presence of a square wave, there should be amplifications at other harmonics (M. Grifoni, private communication) of the fundamental and they should be related to the fundamental SNR as $SNR(\omega)/SNR(n\omega) = n^2$. For a square wave, only the odd harmonics should be present. Figure 4a shows a representative power spectrum from beam 1, plotted in the linear scale to emphasize the peak height. As can be seen, the peak heights scale most closely as $1/n^2$. For a flat background noise spectrum, peak height and SNR are the same quantity. Also, Figures 4b and 4c show the SNRs from the fundamental and the first two odd harmonics. Figure 4b shows all three in the same scale as figure 2b, while figure 4c shows them plotted in a linear scale to emphasize the difference in the SNR. Theoretical investigations in this area are few[21,22] but we hope that with this experiment there may be more work done with this very important distinction.

It is typical that after exceeding the maximum noise needed to induce stochastic resonance, the time scan would be characterised by a very noisy response dominated by incoherent switches caused by the large noise power. In contrast, both of these systems show no switching at all when reaching the upper limit of noise power, instead settling on one state. As mentioned before, this is not a system with a static bistability – rather, its bistable and nonlinear behavior is induced by the presence of a strong drive. This drive and the hysteresis it creates impart a certain amount of rigidity to the system, making switching between the two states difficult. In comparison to the size of the signal required to both form the bistable two-state system and to achieve full switching in the absence of noise (~ 1-10 dBm), the maximum amount of noise induced either externally or by temperature is orders of magnitude smaller. This is certainly a topic worthy of future investigation, as this dynamic bistability is one of the fundamental aspects which departs from the canonical stochastic resonance model.

The presence of stochastic resonance in these systems is exciting not only for its inherent physical interest, but also brings up the possibility of utilizing SR as a means of enhancing signal processing. If the nanomechanical memory element does indeed turn out to be a viable technological innovation, stochastic resonance may very well be one of the mechanisms which allow for its everyday use. Even though our system is manifestly classical, it is closely related to at least one quantum mechanical system (the macroscopic quantum harmonic oscillator[23]). The demonstration of stochastic resonance in this system enables the exploration of stochastic resonance in a quantum mechanical context on a system that is at the very forefront of quantum logic, quantum control, and quantum computation.

We acknowledge the Nanoscale Exploratory Research (NER) program of the National Science Foundation (ECS-0404206) and the DOD/ARL (19-00-2-0004) for the financial support of this research.

**Competing interests statement:** The authors declare that they have no competing financial interests.

**Correspondence and request for materials should be addressed to P.M. (e-mail: mohanty@physics.bu.edu).**

**Figure 1** Schematic of measurement circuit. The circuit serves to excite the nonlinear motion of the beam, modulate the switching, and add in noise to achieve stochastic resonance. The RF source is a Rohde & Schwartz, which operates at a frequency within the bistable region (either 23.4973 MHz or 20.8343 MHz, depending on the beam), at a driving amplitude of 4.0 dBm (beam 1) or 1.0 dBm (beam 2). The modulation is produced by an HP 3325 Synthesizer, with an amplitude of -5.5 dBm (beam 1) or 16.5 dBm (beam 2). In both cases, the modulation is a square wave with a frequency of 0.05 Hz. The white noise source is an HP 33220, capable of creating white noise in a band between 0-15 MHz. For the temperature sweep (beam 2), the noise source was disconnected. The attenuator is an 8 kΩ resistor; the noise source has a separate 30 dB attenuator at the source output. Our beams are 7-8 μm x 300 nm x 200 nm in dimension, and are placed in a $^3$He cryostat (300 mK – 4 K) at the center of a 9T superconducting solenoid magnet.

**Figure 2** Re-emergence of switching behavior as a function of added white noise on beam 1 ($f_{drive}$ = 23.4973 MHz). **a.** The left-hand graphs show selected beam responses as a function of added noise power, with lower powers at the top. The gray bands represent the noise of each state, which includes contributions to the electrical signal from the modulation itself. There is no response until approximately 10 nW of power were introduced (graph A), when a few sporadic switches are seen. As expected, the power spectrum of the response shows no dominant spectral components. With increasing noise power, we observe an increase of switch events, as shown in graphs B (13 nW) and C (27 nW). Finally, as the noise is increased still further, the switches begin to die out (graph D, 32 nW) until we arrive at a regime of no switching (graph E, 63 nW). The right-hand column of plots shows the Fast Fourier Transform (FFT) of the data in the left column. As can be expected, there are sharp peaks at 0.05 Hz and all odd harmonics thereof, as befitting the square-wave response of the switch. **b.** Each individual scan was processed to obtain the signal-to-noise ratio for each noise power. The boldface letters (A, B, C, D, and E) correspond to the respective individual scans shown in Figure 2a.

**Figure 3** Switching behavior as a function of temperature on beam 2 ($f_{drive}$ = 20.8348 MHz). **a.** As the temperature is increased, switching behavior again re-emerges. The modulation is clearly seen in all graphs. Again, the right-hand column shows the FFT of the left-hand column, demonstrating the coherence of the signal as the switches re-emerge. It is important to note that the frequency spectrum shows many more peaks than one would naïvely expect from a simple square-wave response. This is at least partially due to the increased amount of incoherent switches evidenced in this beam. **b.** The SNR was obtained from each individual scan, with a clear resonance occurring around 2600 mK (2.6 K). The line simply connects each individual temperature point, with no fit involved or implied. In contrast to the very broad resonance seen with noise power, this peak is quite sharp and dies off very quickly. This is not unexpected, as the sweep in temperature is much coarser than that used with noise.

**Figure 4** Behaviour of higher harmonics. **a**. The power spectrum from a representative time series, showing the presence of spectral peaks at the odd harmonics of the modulation frequency. This is one of the hallmarks of the square wave drive and response; a sine wave would have a single fundamental peak at the modulation frequency. The three curves represent fits of the peak height versus *n*, the harmonic number. As expected, the peak height goes as $1/n^2$. **b.** Here, we compare the SNR for the higher harmonics, compared to the fundamental. For comparison, all three graphs are plotted in the same scale as Figure 3b. **c.** The SNR for each frequency is plotted again, with the y-axis plotted in a linear, rather than a logarithmic scale, in order to emphasize the difference in SNR for each frequency.

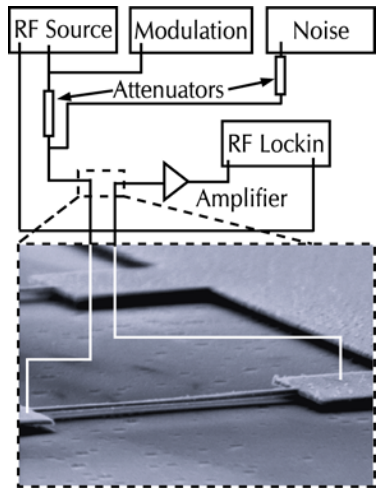

Figure 1

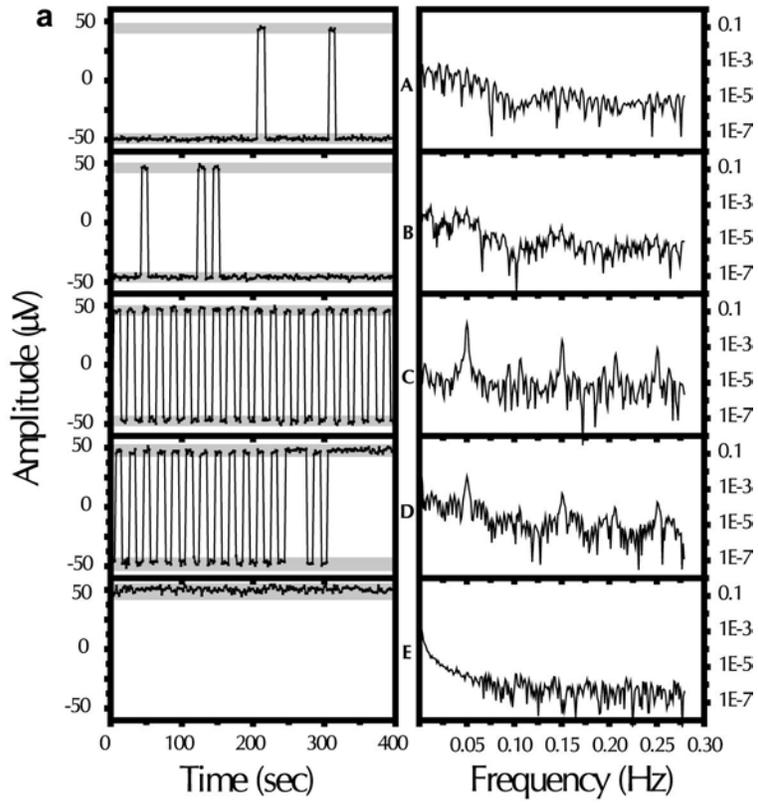
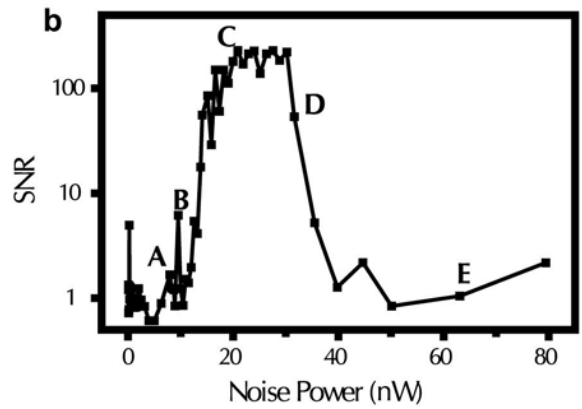

Figure 2

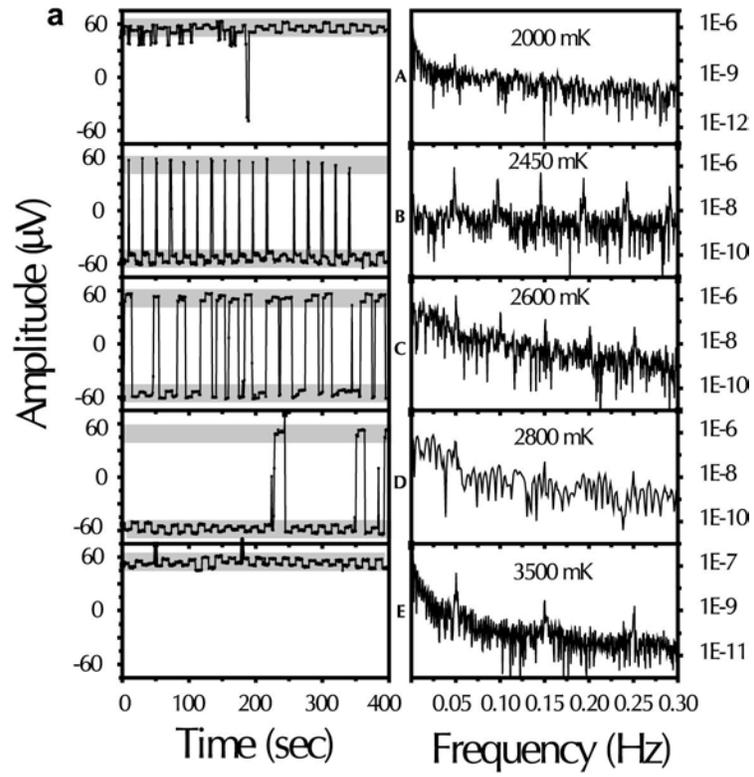
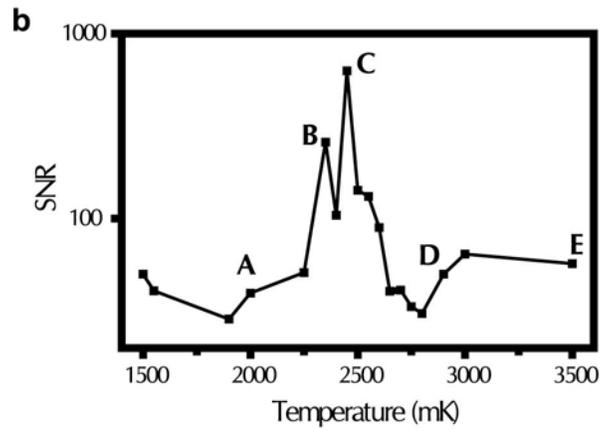

Figure 3

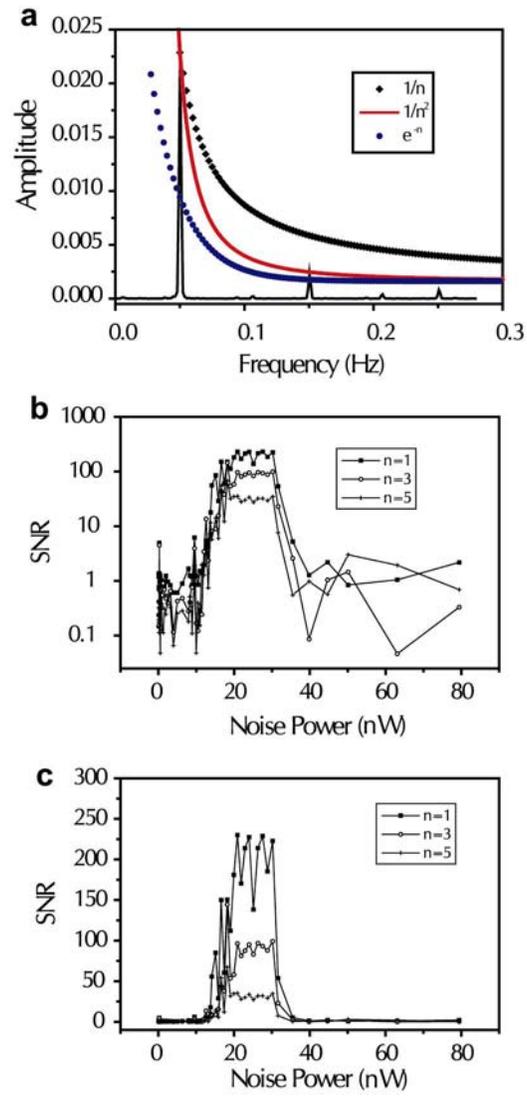

Figure 4